\documentclass[twocolumn,showpacs,amsmath,amssymb]{revtex4}
\usepackage{epsfig}
\usepackage{bm}
\usepackage[usenames]{color}

\begin{document}

\title{On resonance search in dijet events at the LHC}

\author{M. V. Chizhov$^{1,2}$, V. A. Bednyakov$^1$, J. A. Budagov$^1$}
\affiliation{$^{\it 1}$Dzhelepov Laboratory of Nuclear Problems,\\
\mbox{Joint Institute for Nuclear Research, 141980, Dubna,
Russia}\\
$^{\it 2}$Centre for Space Research and Technologies, Faculty of
Physics, Sofia University, 1164 Sofia, Bulgaria}


\begin{abstract}
New strategy for resonance search in dijet events at the LHC is
discussed. The main distribution used  for a bump search is the
dijet invariant mass distribution with appropriated cuts. The
crucial cut, which is applied to maximize signal significance, is on
(pseudo)rapidity difference between the two jets. This is due to the
exponential growing of the QCD background contribution with this
variable. Usually it is assumed that signal from almost all exotic
models populates the central dijet rapidity region $y_{1,2}\simeq 0$
and $|y_1-y_2|\simeq 0$. By contrast, the excited bosons do not
contribute into this region, but produce an excess of dijet events
over the almost flat QCD background in $\chi=\exp|y_1-y_2|$ away
from this region. Therefore, different sets of cuts should be
applied for new physics search depending on the searched resonance
properties. In order to confirm the bump and reveal the resonance
nature various angular distributions should be used in addition. In
particular, for the excited bosons the special choice of parameters
could lead to a dip in the centrality ratio distribution over the
dijet invariant mass instead of a bump, expected in the most exotic
models.
\end{abstract}


\maketitle

\section{Introduction}

Due to the largest cross section of all processes at the hadron
colliders dijet production opens a possibility to search for a
signal, $s$, of new physics in the very early data. In particular, a
bump in the dijet invariant mass distribution would indicate the
presence of a resonance decaying into two energetic partons.
However, due to the huge QCD background, $b$, it is necessary to
optimize signal significance ratio, $s/\sqrt{b}$, in order to
enhance the bump.

The distribution of dijets over the polar angle $\theta$, being
angle between the axis of the jet pair and the beam direction in the
dijet rest frame, is directly sensitive to the dynamics of the
underlying process. While the QCD processes are dominated by
$t$-channel gluon exchanges, which lead to a Rutherford-like
distribution $1/(1-\cos\theta)^2$, exotic physics processes proceed
mainly through the $s$-channel, where the spins of the resonance and
of the initial and final parton states uniquely define the angular
distributions.

The absolute value of the dijet rapidity difference is related to
the polar scattering angle $\theta$ with respect to the beam axis by
$\Delta y\equiv |y_1-y_2|=\ln[(1+|\cos\theta|)/(1-|\cos\theta|)]\ge
0$. Therefore, the QCD background contribution on this variable is
dominated by the exponential growing factor $\exp(\Delta y)$ from
the Rutherford scattering. The choice of the other variable
$\chi\equiv\exp(\Delta y)=(1+|\cos\theta|)/(1-|\cos\theta|)\ge 1$ is
motivated by the fact that Rutherford scattering does not depend on
it and the QCD distribution is almost flat.

In the next section we will consider a model independent signal
distributions of new physics from resonances up to spin of one and
corresponding $\Delta y$ cut optimizations for signal significance.
Present analysis is based on a study of dijet mass angular
distributions~\cite{dijets}, which has been shown to play important
role in disentangling of the resonance properties revealing the
unique signature of the excited bosons.

\section{A unique signal of excited bosons}

Let us consider different possibilities for the spin of a resonance
and its possible interactions with partons. The simplest case of the
resonance production of a (pseudo)scalar particle $h$ with spin 0 in
$s$-channel leads to a uniform decay distribution on the scattering
angle
\begin{equation}\label{G0}
    \frac{{\rm d} \Gamma_0(h\to q\bar{q})}
    {{\rm d} \cos\theta} \propto \vert d^{\,0}_{00}\vert\,^2\sim 1.
\end{equation}

The spin-1/2 fermion resonance, like an excited quark $q^*$, leads
to asymmetric decay distributions for the given spin parton
configurations
\begin{equation}\label{G12}
    \frac{{\rm d} \Gamma_{1/2}(q^*\to qg)}
    {{\rm d} \cos\theta} \propto \vert d^{1/2}_{1/2,1/2}\vert\,^2\sim 1+\cos\theta
\end{equation}
and
\begin{equation}\label{G-12}
    \frac{{\rm d} \Gamma_{1/2}(q^*\to qg)}
    {{\rm d} \cos\theta} \propto \vert d^{1/2}_{1/2,-1/2}\vert\,^2\sim
    1-\cos\theta.
\end{equation}
However, the choice of the variables, which depend on the absolute
value of $\cos\theta$, cancels out the apparent dependence on
$\cos\theta$. In other words, the distributions (\ref{G12}) and
(\ref{G-12}) for dijet events look like uniform distributions
(\ref{G0}).
\begin{figure}[th]
\epsfig{file=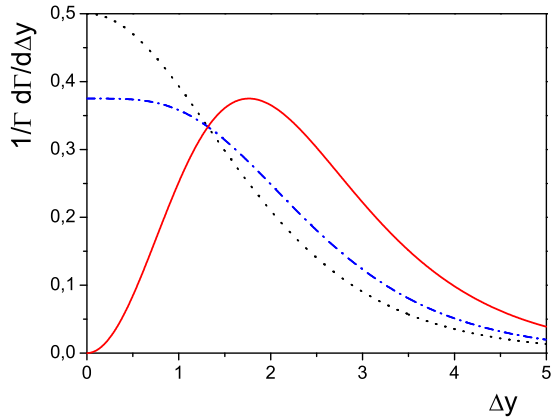,width=0.5\textwidth}
\caption{\label{fig:angular} The normalized angular dijet
distributions as functions of $\Delta y$ for the scalar (or/and
spin-1/2), the gauge bosons with the minimal couplings and the
excited bosons are shown with the dotted, dash-dotted and solid
curves, respectively.}
\end{figure}

According to the simple formula
\begin{equation}\label{trans}
    \frac{{\rm d}\Gamma}{{\rm d}\Delta y}=
    \frac{{\rm d}\cos\theta}{{\rm d}\Delta y}\;
    \frac{{\rm d}\Gamma}{{\rm d}\cos\theta}\,,
\end{equation}
the uniform distribution leads to kinematical peaks at $\Delta y=0$
(the dotted curve in Fig.~\ref{fig:angular})
\begin{equation}\label{dy0}
    \frac{1}{\Gamma_0}\frac{{\rm d}\Gamma_0}{{\rm d}\Delta y}=
    \frac{2{\rm e}^{\Delta y}}{({\rm e}^{\Delta y}+1)^2}.
\end{equation}

According to \cite{dijets,Gia} there are only two different cases
for the spin-1 resonances depending on their interactions with light
fermions.

Usually the gauge bosons $Z'$, $W'$ and axigluons have minimal gauge
interactions with the known light fermions
\begin{equation}\label{LL}
    {\cal L}_{Z'}=\sum_f \left(g^f_{LL}\;\overline{\psi^f_L}\gamma^\mu\psi^f_L
    +g^f_{RR}\;\overline{\psi^f_R}\gamma^\mu\psi^f_R\right) Z'_\mu \,
    ,
\end{equation}
which preserve the fermion chiralities and possess maximal
helicities $\lambda=\pm 1$. At a isospin-symmetric $pp$ collider,
like the LHC, such interactions lead to the specific symmetric
angular distribution of the resonance decay products over the polar
angle $\theta$,
\begin{equation}\label{GLL}
    \frac{{\rm d} \Gamma_1(Z'\to q\bar{q})}
    {{\rm d} \cos\theta} \propto
    \vert d^1_{11}\vert^2+\vert d^1_{-11}\vert^2 \sim
    1+\cos^2\theta \, .
\end{equation}
Similar to the uniform distribution, eq.~(\ref{GLL}) also leads to
kinematical peaks at $\Delta y=0$  (the dash-dotted curve in
Fig.~\ref{fig:angular})
\begin{equation}\label{dy1prime}
    \frac{1}{\Gamma'_1}\frac{{\rm d}\Gamma'_1}{{\rm d}\Delta y}=
    \frac{3{\rm e}^{\Delta y}({\rm e}^{2\Delta y}+1)}{({\rm e}^{\Delta
    y}+1)^4}.
\end{equation}

Another possibility is the resonance production and decay of new
longitudinal spin-1 bosons with helicity $\lambda=0$. The new gauge
bosons with such properties arise in many extensions~\cite{Gia} of
the Standard Model (SM), which solve the Hierarchy Problem. They
transform as doublets $(Z^*\;W^*)$ under the fundamental
representation of the SM $SU(2)_W$ group like the SM Higgs boson
and, therefore, have not minimal gauge interactions with the known
light fermions.

While the $Z'$ bosons with helicities $\lambda=\pm 1$ are produced
in left(right)-handed quark and right(left)-handed antiquark fusion,
the longitudinal $Z^*$ bosons are produced through the anomalous
chiral couplings with the ordinary light fermions
\begin{equation}\label{LR}
    {\cal L}_{Z^*}=\sum_{f}
    \frac{g^f_{RL}}{M}\,\overline{\psi^f_R}\sigma^{\mu\nu}
    \psi^f_L\;\partial_{[\mu} Z^*_{\nu]}+{\rm h.c.}
\end{equation}
in left-handed or right-handed quark-antiquark
fusion~\cite{proposal}. The anomalous interactions (\ref{LR}) are
generated on the level of the quantum loop corrections and can be
considered as effective interactions. The gauge doublets, coupled to
the tensor quark currents, are some types of ``excited'' states as
far as the only orbital angular momentum with $\ell=1$ contributes
to the total angular moment, while the total spin of the system is
zero. This property manifests itself in their derivative couplings
to fermions and a different chiral structure of the interactions in
contrast to the minimal gauge interactions (\ref{LL}).

The anomalous couplings lead to a different angular distribution of
the resonance decay
\begin{equation}\label{GLR}
    \frac{{\rm d} \Gamma_1(Z^*\to q\bar{q})}
    {{\rm d} \cos\theta} \propto
    \vert d^1_{00}\vert\,^2\sim\cos^2\theta
\end{equation}
than the previously considered ones. At first sight, the small
difference between the distributions (\ref{GLL}) and (\ref{GLR})
seems unimportant. However, the absence of the constant term in the
latter case results in novel experimental signatures.

A striking feature of the distribution (\ref{GLR}) is the forbidden
decay direction perpendicular to the boost of the excited boson in
the rest frame of the latter (the Collins--Soper frame~\cite{CS}).
It leads to a profound dip at $\cos\theta=0$ in the Collins--Soper
frame~\cite{proposal} in comparison with the scalar and gauge boson
distributions. The same dips present also at $\Delta
y=0$~\cite{LaTuile10} (the solid curve in Fig.~\ref{fig:angular})
\begin{equation}\label{dy1star}
    \frac{1}{\Gamma^*_1}\frac{{\rm d}\Gamma^*_1}{{\rm d}\Delta y}=
    \frac{6{\rm e}^{\Delta y}({\rm e}^{\Delta y}-1)^2}{({\rm e}^{\Delta
    y}+1)^4}.
\end{equation}

It can be seen from Fig.~\ref{fig:angular} that the excited bosons
have unique signature in the angular distributions. They manifest
themselves through the absolute minimum at $\Delta y=0$ and absolute
maximum at $\Delta y=\ln(3+\sqrt{8})\approx 1.76$\,.

\section{Selection cut optimizations}

From Fig.~\ref{fig:angular} one can be seen that all signal
distributions decrease at $\Delta y\to\infty$, while the QCD
background increases exponentially with $\Delta y$ (as discussed in
the Introduction). Therefore, applying the cut $\Delta y < a$ (i.e.
considering dijet events with $\Delta y < a$ only) one can enhance
the signal significance, ${\cal S} = s/\sqrt{b}$, choosing the
maximum of the corresponding distribution.

So, we can define the relative signal significance as the ratio of
the integrals of the normalized signal distributions (\ref{dy0}),
(\ref{dy1prime}), (\ref{dy1star}) to the QCD background
distribution
\begin{equation}\label{relSig}
    {\cal S}(a) = \frac{\int_0^a \frac{1}{\Gamma}
    \frac{{\rm d}\Gamma}{{\rm d}\Delta y}{\rm d}\Delta y}
    {\int_0^a\exp(\Delta y){\rm d}\Delta y}.
\end{equation}

\begin{figure}[th]
\epsfig{file=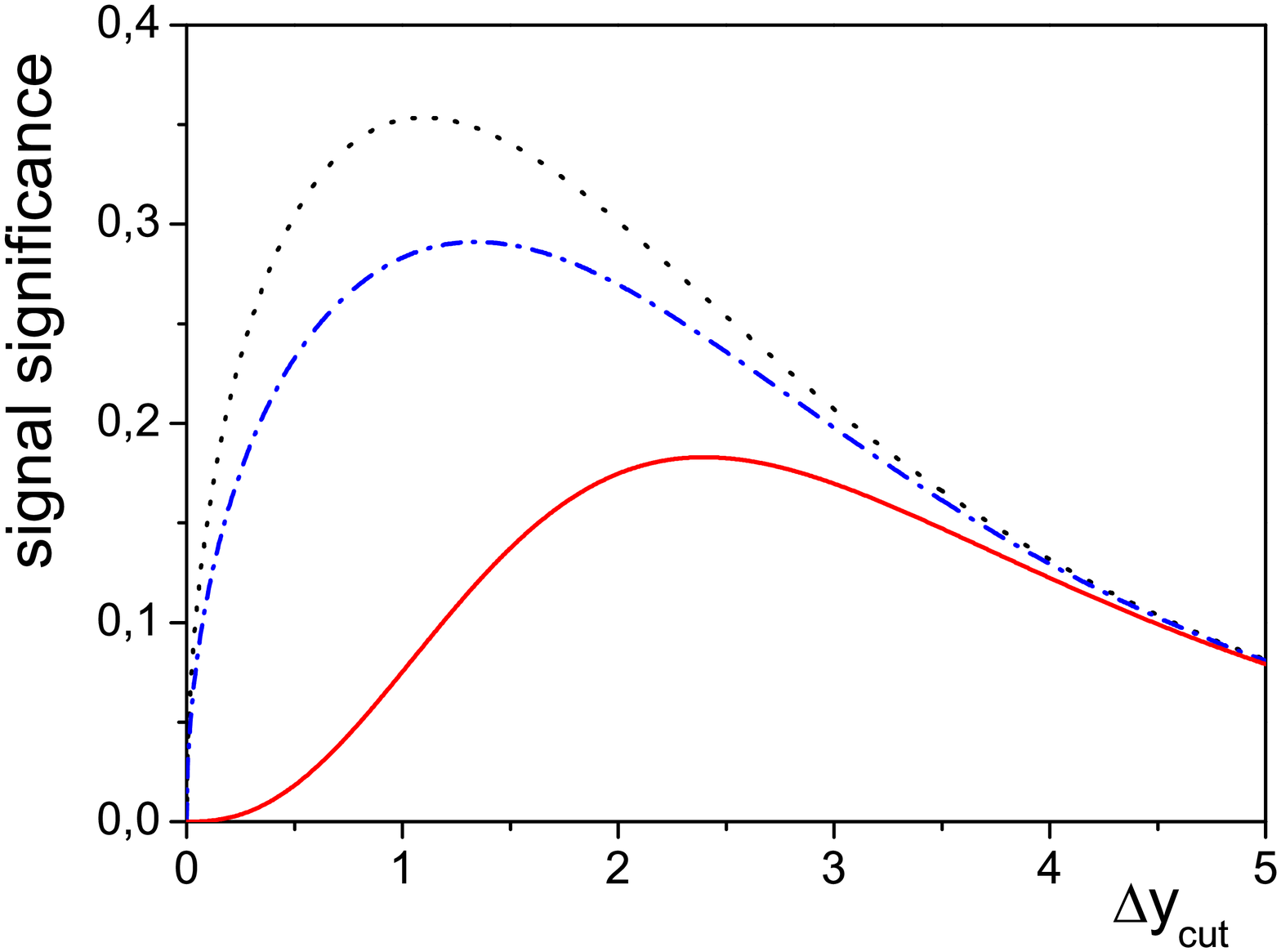,width=0.5\textwidth}
\caption{\label{fig:sig} The relative signal significance
distributions as functions of the cut parameter $a$ for the scalar
(or/and spin-1/2), the gauge bosons with the minimal couplings and
the excited bosons are shown with the dotted, dash-dotted and solid
curves, respectively.}
\end{figure}

Then the significance for the isotropic distribution
\begin{equation}\label{sig0}
    {\cal S}_0(a) = \frac{\sqrt{{\rm e}^a - 1}}{{\rm e}^a + 1}
\end{equation}
reaches the maximum at $a = \ln 3\approx 1.10$ (the dotted curve in
Fig.~\ref{fig:sig}).

The gauge bosons with the minimal coupling (\ref{LL}) possess almost
similar distribution as the scalar and spin-1/2 particles (the
dash-dotted curve in Fig.~\ref{fig:sig})
\begin{equation}\label{sig1prime}
    {\cal S}'_1(a) = \frac{\left({\rm e}^{2a}+{\rm e}^a+1\right)
    \sqrt{{\rm e}^a - 1}}{\left({\rm e}^a + 1\right)^3}
\end{equation}
with the maximum at
$a=\ln(4+\sqrt[3]{764+36\sqrt{401}}/2+20/\sqrt[3]{764+36\sqrt{401}})-\ln3
\approx 1.34$.

These numbers coincide with the selection cut $a=1.3$ applied by the
CMS~\cite{CMS} and ATLAS~\cite{ATLAS} collaborations for the
resonance search in dijet events. However, it is obviously not
optimal for the excited bosons search due to very different shape of
the angular distribution.

Indeed, direct expression of the signal significance function for
the excited bosons
\begin{equation}\label{sig1star}
    {\cal S}^*_1(a) = \frac{\left({\rm e}^a-1\right)^2
    \sqrt{{\rm e}^a - 1}}{\left({\rm e}^a + 1\right)^3}
\end{equation}
lead to the maximum at $a=\ln 11\approx 2.40$ (the solid curve in
Fig.~\ref{fig:sig}). Therefore, application of the usually accepted
cut at $a=1.3$ drastically reduces the number of signal events from
the excited bosons. In other words, there are no model independent
cuts for the resonance search with different spin properties.

Since the peak of the excited boson distribution is shifted from the
origin (Fig.~\ref{fig:angular}), one can expect that applying a more
sophisticated cut $b < \Delta y < a$ the absolute maximum of the
signal significance function of the two variables
\begin{equation}\label{Sig2}
    {\cal S}^*_1(a,b) = \frac{\left(\frac{{\rm e}^a-1}{{\rm
    e}^a+1}\right)^3 - \left(\frac{{\rm e}^b-1}{{\rm
    e}^b+1}\right)^3}{\sqrt{{\rm e}^a-{\rm e}^b}}
\end{equation}
can be found. Precise solution of this problem leads to the
following values of the parameters $a=\ln(6+\sqrt{21})\approx 2.36$
and $b=\ln(6-\sqrt{21})\approx 0.35$ at the maximum. However, an
impact of this solution is expressed in 1.3\% gain in comparison
with the maximum of the expression (\ref{sig1star}).

In order to confirm the event excess in the invariant dijet mass
distribution one can use the various ratios of angular distributions
(see \cite{dijets}). They do not only allow to discover a resonance
but also to reveal its properties. In \cite{ATLAS} new observable
\begin{equation}\label{Fchi}
    F_\chi=\frac{N(\Delta y < b)}{N(\Delta y < a)},~~~~(b < a)
\end{equation}
has been introduced, as the fraction of dijet events produced
centrally versus the events number in more wide region for a
specified dijet mass range. It has been used for quark contact
interaction search at the following parameter values $a=3.4$ and
$b=1.2$\,. Here we will show that the same observable with
appropriated parameters can be used for a resonance search as well.

For example, in the case of the isotropic distribution one can
maximize the deviation from the QCD ratio $F^{\rm QCD}_{\chi}=N_{\rm
QCD}^{(b)}/N_{\rm QCD}^{(a)}$
\begin{eqnarray}
F^0_{\chi}(a,b)&=& \frac{N_{\rm QCD}^{(b)}+N_{\rm new}^{(b)}}{N_{\rm
QCD}^{(a)}+N_{\rm new}^{(a)}}
\nonumber \\
&\approx& F^{\rm QCD}_{\chi}
    \left(1+\frac{N_{\rm new}^{(b)}}{N_{\rm QCD}^{(b)}}
           -\frac{N_{\rm new}^{(a)}}{N_{\rm QCD}^{(a)}}
    \right)
\nonumber \\
    &\equiv& F^{\rm QCD}_{\chi}+\delta F^0_\chi > F^{\rm QCD}_{\chi}.\label{minEta}
\end{eqnarray}
Since $N_{\rm QCD}^{(a)}\propto{\rm e}^{a}-1$ and $N_{\rm
new}^{(a)}\propto ({\rm e}^{a}-1)/({\rm e}^{a}+1)$ the absolute
maximum appears at $a=\log(2+\sqrt{5})\approx 1.44$ and
$b=\log(\sqrt{5})\approx 0.80$\,.

For the spin-1 resonances the expressions $N^{(a)}_{\rm new}$ are
more complicated and solutions cannot be expressed through radicals.
So, for the gauge bosons with the minimal interactions $N^{(a)}_{\rm
new}\propto ({\rm e}^{3a}-1)/({\rm e}^{a}+1)^3$ and the absolute
maximum is reached at $a\approx 1.63$ and $b\approx 0.94$\,. For the
both considered cases there are only positive contributions to
$F_\chi$ from the new physics and $F_\chi$ is always greater than
$F^{\rm QCD}_\chi$.

\begin{figure}[th]
\epsfig{file=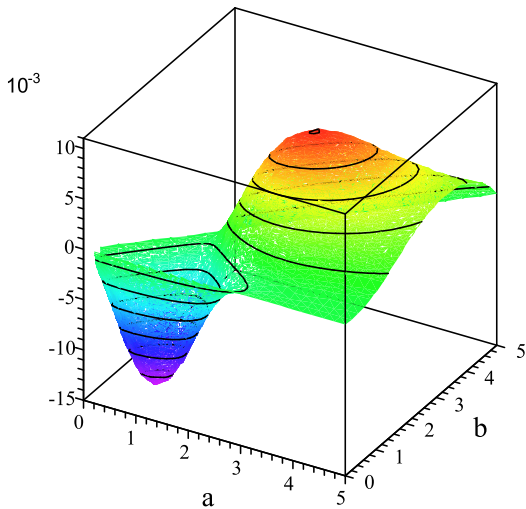,width=0.45\textwidth}
\caption{\label{fig:shape} The $F^{*1}_\chi$ shape for the excited
bosons.}
\end{figure}

By contrast to this, the excited bosons represent more interesting
case with absolute minimum and maximum at $a\approx 1.07$, $b\approx
0.57$ and $a\approx 3.10$, $b\approx 2.25$, correspondingly
(Fig.~\ref{fig:shape}),
due to a peculiar shape of the angular distribution $N^{(a)}_{\rm
new}\propto ({\rm e}^{a}-1)^3/({\rm e}^{a}+1)^3$.

\section{Discussions and comparisons}

Since the scalar resonances peak at origin more stronger than spin-1
gauge bosons with the minimal couplings and the maximum of the
excited bosons distribution is shifted away from the origin, they
have decreasing relative signal sensitivities at their maxima
\begin{equation}\label{propSig}
    \left.{\cal S}_0\right|_{\rm max}:\left.{\cal S}'_1\right|_{\rm max}
    :\left.{\cal S}^*_1\right|_{\rm max} \simeq 1.2:1.0:0.6\,.
\end{equation}
Here we assume the same production cross section for all resonances.
Moreover, if the selection cut $\Delta y < 1.3$ is applied for the
case of the excited bosons search, it reduces 1.6 times the
sensitivity than the optimal cut $\Delta y < 2.4$\,. For example, if
a some insignificant 2$\sigma$ excess is observed for the commonly
accepted cut, it could become an evidence for the optimal cut in the
excited bosons case.

In order to obtain absolute signal significances for some models we
have used the CalcHEP package~\cite{CalcHEP}. Let us compare two
models~\cite{LaTuile10} for the charged spin-1 resonances $W'$ and
$W^*$ with mass of 1 TeV, which have the same production cross
sections at the LHC in the leading order approximation. The signal
events and QCD background have been generated with MRST2007lomod
PDFs at 7~TeV center-of-mass energy and the following jet cuts:
$M_{jj} > 700$~GeV and $|\eta_j| < 2.5$\,.

In order to optimize search the additional selection cut $\Delta y <
1.3$ has been applied in the case of $W'$ resonances and $\Delta y <
2.4$ for the $W^*$ case. So, already 100~pb$^{-1}$ is enough to
obtain 5$\sigma$ signal significance in the first case and
230~pb$^{-1}$ for the second case. If we have applied $\Delta y <
1.3$ cut for the excited boson $W^*$, 560~pb$^{-1}$ will be
necessary to reach the same sensitivity.

It should be stressed also, that the ratio $F_\chi$ is less affected
by the systematic errors. However, the impact of new physics from
the different resonances to $F_\chi$ occurs as
\begin{equation}\label{propF}
    \left.\delta F^0_\chi\right|_{\rm max}:\left.\delta F'^1_\chi\right|_{\rm max}:
    \left.\delta F^{*1}_\chi\right|_{\rm min}\simeq 1.4:1.0:-0.4\,.
\end{equation}
Therefore, the contribution of spin-1 resonances and mainly the
excited bosons are more suppressed than in (\ref{propSig}).
Nevertheless, for the excited bosons a unique possibility exists
that the absolute minimum and maximum are present. In these cases we
have $\left.F^{*1}_\chi\right|_{\rm min} < F^{\rm QCD}_\chi$ and
$\left.F^{*1}_\chi\right|_{\rm max} > F^{\rm QCD}_\chi$,
correspondingly. Although the absolute minimum is 1.3 times dipper
than the maximum value, more wider regions in the later case lead to
smaller uncertainties. It will allow to verified once again the
excited bosons signature.


In conclusion we would like to note that in this paper we have
considered the novel optimization of angular cuts aimed at the most
effective search for the new resonances with different angular
distributions of their decay products. Simple analytical formulas
have been obtained. Even for the same resonance spin there are
different optimized cuts with respect to the resonance interactions.

\section*{Acknowledgements}
The work of M.V. Chizhov was partially supported by the grant of
Plenipotentiary of the Republic of Bulgaria in JINR for 2011 year.
He is also acknowledged to a discussion with Prof. Tejinder Virdee.


\end{document}